\documentclass[12pt,preprint]{aastex}

\begin{document}

\title{On interpretation of recent proper motion data
for the Large Magellanic Cloud}

\author{Kenji Bekki} 
\affil{
ICRAR,
M468,
The University of Western Australia
35 Stirling Highway, Crawley
Western Australia, 6009, Australia
}

\begin{abstract}

Recent observational studies using
the {\it Hubble Space Telescope (HST)} have derived 
the center-of-mass proper motion (CMPM)
of the Large Magellanic Cloud (LMC). 
Although these studies  carefully treated
both rotation and perspective effects in deriving the proper motion
for each of the sampled fields, 
they did not consider the effects of local random motion
in the derivation.
This means that
the average PM of the fields (i.e., the observed CMPM)
could significantly deviate from the true CMPM,
because the effect of local random motion can not be close to zero
in making the  average PM for the small number of the fields ($\sim 10$).
We discuss how significantly the observationally derived CMPM
can deviate from the true CMPM by applying
the same method as used in the  observations for
a dynamical model of the LMC  with a known true CMPM.
We find that the deviation can be as large as $\sim 50$ km s$^{-1}$
($\sim 0.21$ mas yr$^{-1}$), if the LMC has a thick disk and
a maximum circular velocity of $\sim 120$ km s$^{-1}$.
We also find that 
the deviation depends both  on the total number of the sampled
fields and on structure and kinematics  of the LMC.
We therefore suggest that there is a possibility that the observed CMPM of the LMC 
deviates from the true one to some extent.
We also show that a simple mean of PM for a large number of the LMC fields ($\sim 1000$)
can be much closer to the true CMPM.
\end{abstract}


\section{Introduction}

The orbital evolution of the Magellanic Clouds (MCs) with respect to
the Galaxy has been considered to be  one of key parameters that control 
dynamical and chemical evolution of the LMC and the Small Magellanic
Cloud (SMC) and the formation processes of the Magellanic Stream (MS)
and its leading arms (e.g., Murai \& Fujimoto 1980, MF80; Gardiner \& Noguchi 1996,
GN96;
Bekki \& Chiba 2005, BC05; Mastropietro et al. 2005).
Recent observational studies on the CMPM for  the LMC
using the High Resolution Camera (HRC) of the
Advanced Camera for Surveys (ACS) on the $HST$ 
have derived an  accurate present 3D motion of the LMC around the Galaxy 
(e.g., Kallivayalil et al. 2006, K06a; Piatek et al. 2008, P08).
One of remarkable results from these studies
is that the LMC can possibly have a  velocity 
with respect to the Galaxy ($V_{\rm LMC}$) being  $\sim 380$ km s$^{-1}$
(Kallivayalil et al. 2006),
which is significantly larger than that ($V_{\rm LMC} \sim 300$ km s$^{-1}$)
predicted by one of promising  theoretical
models for the formation of the MS (e.g., GN96).
The observational results thus have profound implications on 
the past orbits of the MCs 
(e.g., Besla et al. 2007; Kallivayalil et al. 2006, K06b),
a possible common halo of the MCs (Bekki 2008), 
and the formation processes of the MS (e.g., Diaz \& Bekki 2011, DB11;
Ruzicka et al. 2010, R10).

The previous work by K06a considered that the observed PM for each field
(PM(field)) 
is a combination of the PM  of the center-of-mass (CM) for the
LMC (PM(CM)) and the field-dependent residual 
(PM$_{\rm res}$(field)) as follows:
\begin{equation}
{\rm PM(field)=PM(CM)}+{\rm PM}_{\rm res}({\rm field}) .
\end{equation}
In estimating  the PM of the LMC CM (i.e., PM$_{\rm est}$(CM))  for 
each of the selected high-quality 13 fields,
K06a very carefully considered how
the internal rotation of the LMC (``rotation effect'') 
and the viewing angle (``perspective effect'') influence
PM$_{\rm res}$(field) and thereby
made an average of the 13 PM$_{\rm est}$(CM) to derive the CMPM.
Since the average PM is not a {\it simple} average of the observed
PM of the 13 fields (i.e., not the average of PM(field)),
the observed CMPM can be pretty close to the true one,
if all stars have almost circular motion and if the LMC has a very thin disk.
However, the LMC has a thick disk with a bar 
(e.g.,van der Marel et al. 2002, vdM02),
which is indicative of larger local random motion both in radial and vertical
directions (i.e., deviation from circular motion). 
Therefore, PM$_{\rm est}$(CM) for each field can
significantly deviate from the true PM of the LMC  and 
the average PM$_{\rm est}$(CM) can  
also deviate from the true CMPM if the number of the sampled field is small.

The purpose of this Letter is to show how significantly the observationally
derived CMPM can deviate from the true one by using a dynamical (N-body)  model
for the LMC with a known true CMPM. 
In the present study,
we first pick up randomly stellar particles with
the particle number ($N$) of $3-3000$ in a N-body model
for the LMC with a given structure and kinematics  and thereby
derive the CMPM of the LMC in the same way as  done in 
previous observational studies.
We then compare the derived CMPM with the true one so that we can
discuss the possible difference between the two CMPMs.
This investigation is quite important, because
the possible difference between the observed and true CMPMs can not be 
observationally discussed  owing to the lack of detailed information
of the 3D positions and velocities  of stars in each field.

Recent different observational studies on the CMPM of the LMC have revealed different
CMPM and maximum circular
velocity  ($V_{\rm c}$) of the LMC
(e.g., K06a, P08, and Costa et al. 2009; C09), and
different PM studies using almost the same data set and adopting a similar
PM measurement method have  derived different CMPMs and $V_{\rm c}$ of the LMC
(K06a and P08): this is yet to be explained.
Furthermore P08 already suggested that a significant scatter ($\sim 30$ km s$^{-1}$)
in the derived PM of the sampled 21 LMC fields
is due to significant departures from circular
motion.  Thus it is crucial to investigate how random motion in the LMC
can affect the observational estimation of the CMPM in a quantitative way.

\section{The model}

\subsection{The LMC}

The present LMC model is consistent with
a high-mass model in BC05 in terms
of the disk structure and the  dark matter density profile,
but it is slightly different from BC05 in the  dark matter fraction
and the inner rotation curve profile (also the absence of gas).
The modeled LMC is consistent with
the observed radial structure of the disk
(e.g., Bothun \& Thompson 1988),  the total mass
(e.g., Westerlund 1997; P08), structure and kinematics of 
the thick disk (vdM02), and dark matter content (vdM02). 
The LMC is composed of a  dark matter halo
and a stellar disk
with the total masses being  $M_{\rm dm}$ and $M_{\rm d}$, respectively.
Following the observational results by vdM02 
showing  $M_{\rm dm} = (8.7 \pm  4.3) \times 10^9 M_{\odot}$  within 9kpc
of the LMC,
we assume that a reasonable mass fraction of the dark matter halo
($f_{\rm dm}=M_{\rm dm}/(M_{\rm dm}+M_{\rm d}$)) 
is 0.50-0.67 within the adopted LMC size.
We adopted an NFW halo density distribution (Navarro, Frenk \& White 1996)
suggested from CDM simulations and the ``c''-parameter is set to be 12.
The dark matter halo is truncated at the observationally suggested tidal
radius of the LMC ($\sim 15$ kpc; vdM02).  
We mainly investigate the ``fiducial'' 
LMC model with 
the total mass ($M_{\rm t}=M_{\rm dm}+M_{\rm d}$)
of $2 \times 10^{10} M_{\odot}$, $f_{\rm dm}=0.5$, and  $V_{\rm c}=117$ km s$^{-1}$.

The radial ($R$) and vertical ($Z$) density profiles of the disk
(with the size $R_{\rm d}$ of 7.5 kpc) were
assumed to be proportional to $\exp (-R/R_{d,0})$, with scale
length $R_{d,0}$ = 0.2$R_{\rm d}$, and ${\rm sech}^2 (Z/Z_{d,0})$, with scale
length $Z_{d,0}$ = 0.06$R_{\rm d}$,  respectively:
the stellar disk has the radial and vertical scale length of 1.5 kpc and 0.45 kpc,
respectively.
In addition to the rotational velocity caused both by the gravitational fields
of
the disk and dark halo components,
the initial radial and azimuthal velocity dispersions
were  assigned to the disk
component according to the epicyclic theory with Toomre's parameter $Q$ 
(Binney \& Tremaine 1987) ranging
from 0 to 3 in the present study. By investigating models with different $Q$,
we can discuss how local random motion (due to non-zero velocity dispersions) 
can introduce differences
between the observed and true CMPMs.
We run  the LMC disk model for 20 dynamical time scales
so that we can construct a ``barred model'' for our investigation.
There is no significant differences in stellar kinematics between
the above relaxed and unrelaxed models.
The line-of-sight velocity dispersions
in the $x$-, $y$-, and $z$-components of stellar velocities
(${\sigma}_{x}$, ${\sigma}_{y}$, and ${\sigma}_{z}$, respectively)
reach their maximum values
of 55 km s$^{-1}$, 55 km s$^{-1}$, and 43 km s$^{-1}$, respectively, 
at $R=0$ in the standard LMC model with $Q=1.5$.

\subsection{Estimation of possible differences between simulated and true 
center-of-mass velocities}

We investigate  possible differences between observed and true 3D velocities
of the CM of the LMC 
rather than the CMPM differences, because we can more clearly show
the differences without considering the coordinate transformation and the present
location of the LMC with respect to the Galaxy.
The CM  of the 
LMC is assumed to be
located at the center of the coordinate (i.e, ($x$, $y$, $z$) = (0, 0, 0))
in all models.
The spin of the LMC disk 
is specified by two angles $\theta$ and
$\phi$ in units of degrees, where
$\theta$ is the angle between
the $z$-axis and the vector of the angular momentum of the disk,
and $\phi$ is the azimuthal angle measured from $x$ axis to
the projection of the angular momentum vector of the disk onto
the $x$-$y$ plane.

The LMC in a model is assumed to be moving
with  a ```true'' 3D CM velocity  
of {\bf  V} = ($V_{\rm x}$, $V_{\rm y}$, $V_{\rm z}$) and therefore 
the initial velocity  of each $i$-th stellar particle (${\bf v}_{i}$) within the LMC
is described as: 
\begin{equation}
{\bf v}_i={\bf V} + {\bf  V}_{i},
\end{equation}
where {\bf V}$_{i}$ is the 3D motion of the particle  with respect to
the center of mass of the LMC and depends on the location of the particle  and the adopted
dynamical model of the LMC (e.g., $V_{\rm c}$).

We first randomly pick up stellar particles with the total number of $N$
from 200000 particles in the LMC model. Each selected stellar particle
is not literally a star but represents a field with local kinematics. 
Then we derive the 3D velocity
of the CM ({\bf v}$_{\rm cm, \it i}$) from that of each field 
as follows:
\begin{equation}
{\bf v}_{\rm cm, \it i}={\bf  v}_{i}-{\bf V}_{\rm c, \it i},
\end{equation}
where {\bf V}$_{\rm c, \it i}$ is the 3D circular velocity vector at
the position of the field  and thus depends on inclination  angles
(i.e., $\theta$ and $\phi$)  of the LMC
and dynamical properties of the LMC (e.g., $V_{\rm c}$): thus we here
consider both rotation and perspective effects properly.
This derivation of {\bf v}$_{\rm cm, \it i}$ is done in 
the same way as done in the previous {\it HST} PM studies (e.g., K06a).
Unlike observations,  there is no uncertainty in {\bf V}$_{\rm c, \it i}$,
because both the radial dependence of the rotation curve
and inclination angles (corresponding to viewing angles in
observations) in the LMC are precisely considered.

Thus the simulated CM 3D velocity ({\bf V}$_{\rm sim}$) is described as 
follows:
\begin{equation}
{\bf V}_{\rm sim}= \frac{1}{N} \sum_{i=1}^N {\bf v}_{\rm cm, \it i}.
\end{equation}
The above {\bf v}$_{\rm cm, \it i}$ can deviate significantly from
the true  {\bf V} because of velocity dispersions of the LMC.
We estimate the deviation for the sampled $N$ fields 
in a quantitative way as follows:
\begin{equation}
\Delta {\bf V}_{\rm m}= {\bf V}_{\rm sim} -  {\bf V}.
\end{equation}
We also derive  the difference between
the simulated and true 3D velocities ($\Delta {\bf V}_{i}$)
for each field
as follows:
\begin{equation}
\Delta {\bf V}_{i}= {\bf v}_{\rm cm, \it i}-{\bf  V}.
\end{equation}
We estimate  $\Delta {\bf V}_{\rm m}$ 
(=($\Delta{\rm V}_{\rm m, x}$,
$\Delta{\rm V}_{\rm m, y}$,
$\Delta{\rm V}_{\rm m, z}$)) and the absolute magnitude
of the projected velocity 
($\Delta{\rm V}_{\rm m}=\sqrt{ \Delta{\rm V}_{\rm m, x}^2 
+ \Delta{\rm V}_{\rm m, y}^2 }$) for each sample.
The method used for deriving {\bf V}$_{\rm sim}$  by  the above
equations (3)-(6) 
is referred to as ``the standard'' method.

We also adopt a ``simple method'' in which 
${\bf V}_{\rm sim}$ is derived without considering rotation and perspective effects:
we estimate  ${\bf V}_{\rm sim}$ simply by making an average of
${\bf  v}_{i}$ (i.e., ignoring ${\bf V}_{\rm c, \it i}$  in the equation (3)).
By comparing the results derived by  the simple and standard methods,
we can demonstrate that it is important 
to consider rotation and perspective effects in deriving the CMPM of the LMC.
The present results do not depend on {\bf V} at all, and thus  we show the results
for the models with {\bf V}=(0, 0, 0) for clarity.
We mainly show the results for the standard LMC model with $\theta =45^{\circ}$ 
and $\phi = 30^{\circ}$ in the present study,
because the results of other models are essentially the same as those in this
standard model.

\section{Results}

Fig. 1 shows that 
the difference between the velocity of each field  and true CM velocity
can be quite large and become up to $\sim 100$ km s$^{-1}$ in each 
velocity component for some fields, if the simple method
(i.e., ${\bf v}_{\rm cm, \it i}-{\bf  V} = {\bf v}_i- {\bf V}$)
is applied for this fiducial  LMC model with $V_{\rm c}=117 $ km s$^{-1}$.
Also the simulated CM velocity for the 13 fields
({\bf V}$_{\rm sim}$)
can not be close to the true one owing to the local random motion
($\Delta {\rm V}_{\rm m} \sim 33$ km s$^{-1}$).
On the other hand,
although the difference between the velocity of each field  and 
the true CM velocity
can be large (up to $\sim 50$ km s$^{-1}$) 
in the standard method, 
the simulated LMC CM velocity ({\bf V}$_{\rm sim}$)
can be pretty close to the true one 
($\Delta {\rm V}_{\rm m} \sim 8$ km s$^{-1}$).
This clearly demonstrates that if rotation and perspective effects
are taken into account,
the simulated CM velocity can be closer to the true one.

Fig. 2 shows how the deviation from  the simulated LMC CM velocity
({\bf V}$_{\rm sim}$)
from the true CM one 
depends on $Q$ and presence or absence of the central stellar bar in the LMC
disk. The simulated CM velocities are derived for 1000 different data samples 
each of which has $N$ fields.
The difference  between the simulated and true velocities can be as large
as 50 km s$^{-1}$ in the non-barred model with $Q=1.5$,
though the average  $\Delta {\rm V}_{\rm m}$ for the 1000 samples  is only
$\sim22$ km s$^{-1}$.  The origin of the significant deviation in some samples 
is due to local random motion in the LMC disk.
It is also clear that the simulated CM velocity can deviate more significantly
from the true one in the model with a higher $Q$ owing to the larger degree of
local random motion:
the average  $\Delta {\rm V}_{\rm m}$ 
is $\sim 35$  km s$^{-1}$ for the non-barred model
with $Q=3.0$.
The average $\Delta {\rm V}_{\rm m}$ is $\sim  25$ km s$^{-1}$ for the barred model,
which means that
there is no remarkable difference between non-barred and barred models

Fig. 3 shows the degree of deviation of the simulated LMC CM velocity from
the true one depends strongly on the number of fields ($N$) 
such that it can be  higher in models with smaller $N$:
the average  $\Delta {\rm V}_{\rm m}$ for $N=3$, 30, and 300
are 40 km s$^{-1}$, 12 km s$^{-1}$, and 4 km s$^{-1}$, respectively. 
It should be stressed  here that even if 30 fields are used,
the deviation can be still as large as 30 km s$^{-1}$ in each velocity component.
Fig. 3 also shows that the deviation of the simulated CM velocity
can be quite small ($<10$ km s$^{-1}$), if 300 field are used. 
This result implies that at least $\sim 300$ fields are necessary
to make a very accurate CMPM of the LMC if we estimate the CMPM
using the method by K06a. Fig. 4 demonstrates that 
even if we adopt the simple method,  the difference between the simulated
and true CM velocities can be well less than 10 km s$^{-1}$ for 
$N>300$.
This result implies that the simple average of the observed PMs for a large
number of fields (or stars) can give a quite accurate CMPM of the LMC.

The present results depend on model parameters as follows:
The deviation of the simulated LMC CM velocity from the true one 
can be smaller 
in the low-mass LMC model with $M_{\rm t}=10^{10} M_{\odot}$,
$f_{\rm dm}=0.5$, and $V_{\rm c}=83$ km s$^{-1}$:
the mean ${\rm V}_{\rm m}$ 
is 16 km s$^{-1}$.
The present results do not depend strongly on $f_{\rm dm}$: 
the mean ${\rm V}_{\rm m}$  
for $f_{\rm dm}=0.5$ and 0.67 are  22 km s$^{-1}$ and 18 km s$^{-1}$, respectively.
The deviation of the simulated CM velocity of the LMC  from the true one 
can be larger  
in the models with higher inclination angles; for example,
the mean ${\rm V}_{\rm m}$ is 34 km s$^{-1}$ for $\theta=60^{\circ}$ and 
$\phi = 80^{\circ}$.

\section{Discussion and conclusions}

The present study has first demonstrated that the LMC CMPM derived from 
$\sim 10-20$ fields 
can possibly deviate significantly from the true one 
(up to $\Delta V_{\rm m} \sim 50$ km s$^{-1}$)
even if rotation and perspective effects are carefully taken into account.
Also the deviation of the  observed CMPM from the true one  depends on structural
and kinematical properties (e.g., $V_{\rm c}$) of the LMC.
The present study however can  not claim 
that the 3D motion of the LMC by K06a deviates significantly  from the true one,
but it strongly suggests that there is a possibility that the observed
3D motion using the PMs for a small number of the LMC fields
{\it can} be significantly different from the true one owing to 
local random motions of stars in the LMC.

The possible deviation is very significant in terms of modeling the MS,
because only a $\sim 10$ km s$^{-1}$ difference in a velocity component 
can cause a significant difference in the orbital evolution of the MCs
(e.g., the longevity of the MCs' binary status)
that is crucial for the formation process of the MS
(e.g., MF80). 
Our study also has pointed out that  $\sim 300$ fields of the LMC
are necessary to derive a  quite accurate CMPM for the LMC,  because the effect of 
local random motion in deriving the CMPM can become negligible for
such a large number of the fields.
The present study suggests that
the CMPM of the SMC derived by K06b  can be  less accurate than that of the LMC,
firstly because
only a  small number of the SMC fields (5) are  used for the PM measurement,
and secondly because the stellar component can be better modeled
as a system dynamically supported
by velocity dispersion (e.g., Bekki \& Chiba 2009).

The latest PM study of the LMC by Vieira et al. (2010, V10)
has derived the CMPM  by simply making an average of the observed PMs for
3822 stars and thereby
 has shown that the CMPM is  different
from those derived in previous studies (e.g., K06a,b; P08);
V10 have also shown  that
the LMC  is currently orbiting the Galaxy  with 
$V_{\rm LMC}=343 \pm47.8$ km s$^{-1}$, which is significantly 
smaller than $V_{\rm LMC} \sim 380$ km s$^{-1}$ derived by
K06a using the $HST$ proper motion results.
Our results suggest that the simple average for 3822 stars in V10
can give an accurate CMPM,  if observation-related errors for
the PM measurement are really small for each field (star).
This  suggests that the current 3D velocity of
the LMC by V10 should be also properly included in
modeling the MCs, in particular,  the formation of the MS.

If $V_{\rm LMC}$ by V10 is really close to the true one, 
then the results by V10 would have  a profound implication for the dynamical
modeling of the MS formation.
Recent dynamical models that are consistent with the CMPM by K06a
(i.e., higher $V_{\rm LMC}$)
have failed to reproduce the bifurcated structure of the MS
 and the elongated  leading arms  self-consistently
(e.g., Besla et al. 2010; R10).
DB11 have successfully reproduced the above two fundamental
observations  in a model with bound orbits of the MCs
(with respect to the Galaxy) yet
a rather high circular velocity (250 km s$^{-1}$) of the Galaxy.
They thus  suggested that bound orbits of the MCs are  necessary to reproduce
the two observations.

If the present 3D motion of the LMC by V10 is pretty close to the true one,
then the LMC 
is  bound to the Galaxy with the maximum circular
velocity of  $220-250$ km s$^{-1}$. This means that the MS can be formed as
a result of the LMC-SMC-Galaxy interaction, as shown by previous MS formation  models
with lower $V_{\rm LMC}$ (e.g., GN96). 
Given different values of the present 3D velocities of the LMC derived
by different observations (e.g., K06a, P08, C09, and V10),
future theoretical studies  for the MS formation would need to investigate
at least all of the three representative models with $V_{\rm LMC}$ 
$ 300$ km s$^{-1}$ (C09),
$ 340$ km s$^{-1}$ (V10),
and $ 380$ km s$^{-1}$ (K06a).
In this situation,  it would be fair to claim that a theoretical MS formation
model that can  best reproduce fundamental properties of the MS and the leading arms 
would  predict the LMC CMPM that is closest to the true one.

\acknowledgments
I am  grateful to the anonymous referee for constructive and
useful comments.

\begin{figure}
\epsscale{1.0}
\plotone{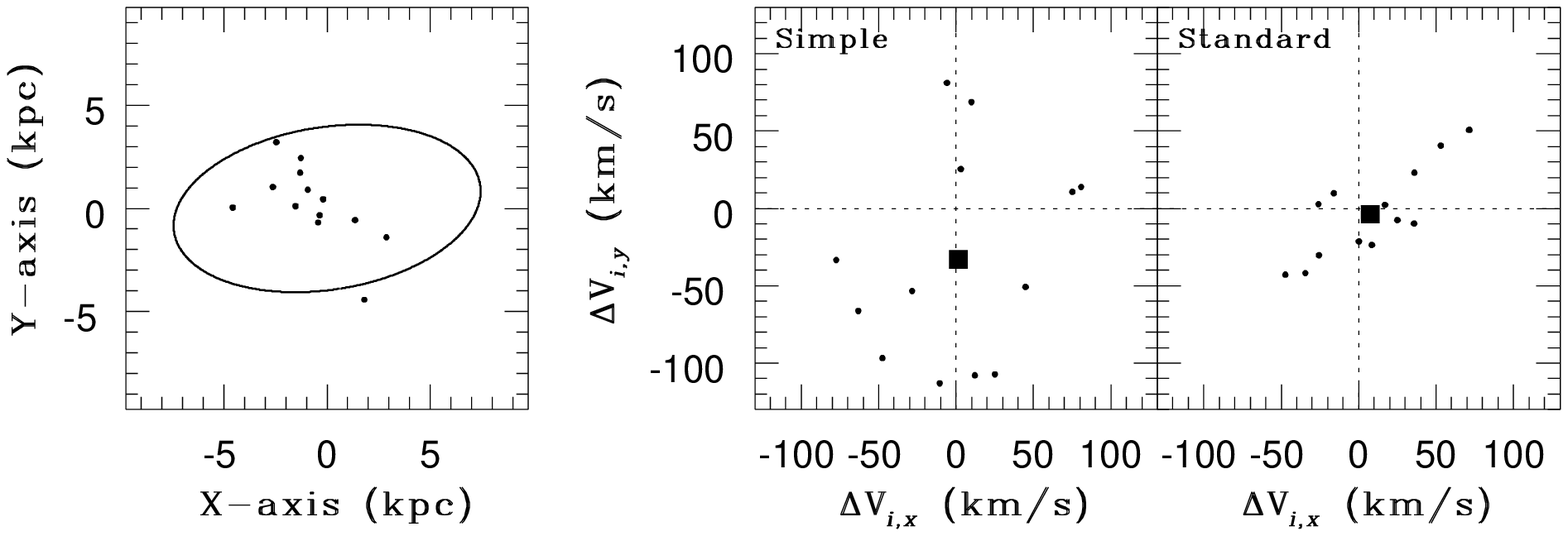}
\figcaption{
The left panel shows 
the spatial distribution 
of the selected 13 fields 
projected onto the $x$-$y$ plane
in the fiducial LMC model with
$Q=0$ 
and inclination angles $\theta$ and $\phi$ being $45^{\circ}$ and $30^{\circ}$,
respectively.  The  elongated circle describes the outer edge of the LMC disk
($R_{\rm d}=7.5$ kpc).
In the right two panels,
each small dot describes the deviation of the simulated CM velocity of
each field from the true
one ($\Delta {\bf V}_i={\bf v}_{\rm cm, \it i}-{\bf V}_i$)
derived by the simple method (left) and by the standard one (right).
Here the $x$- and $y$-components of $\Delta {\bf V}_i$ 
($\Delta V_{i, \rm x}$ and $\Delta V_{i, \rm y}$, respectively) are shown.
The big filled square represents the average of  $\Delta {\bf V}_i$ 
and thus describes the deviation of the  simulated
CM velocity of the LMC from the true one.
The simulated LMC CM velocity 
can deviate from the true one even in the model
owing to the thick disk with non-zero
vertical velocity dispersion.
\label{fig-1}}
\end{figure}

\begin{figure}
\epsscale{1.0}
\plotone{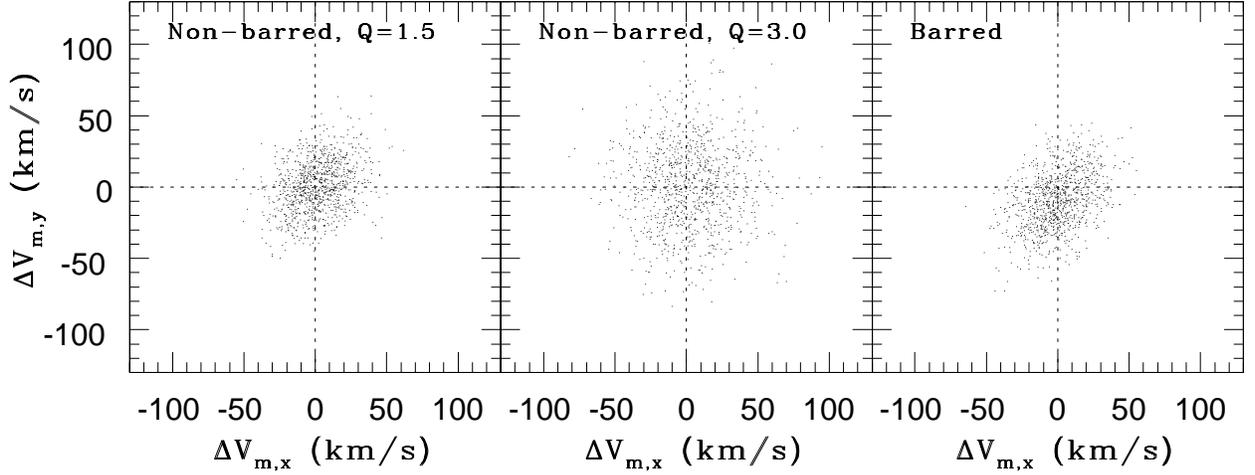}
\figcaption{
The plot of $\Delta V_{\rm m,y}$ as a function of $\Delta V_{\rm m,x}$
for each data sample
in the fiducial LMC model with $Q=1.5$ and no bar (left),
$Q=3.0$ and no bar (middle), and a central stellar bar (right).
10 fields are used 
in deriving 
$\Delta {\bf V}_{\rm m}$
for each sample  and the results of 1000 samples are shown in each frame. 
It is clear that the simulated CM velocity of the LMC can significantly deviate
from the true one owing to local random motion (i.e., non-zero 
velocity dispersion),
in particular, in the model with a higher degree of random motion ($Q=3.0$).
\label{fig-2}}
\end{figure}

\newpage

\begin{figure}
\epsscale{1.0}
\plotone{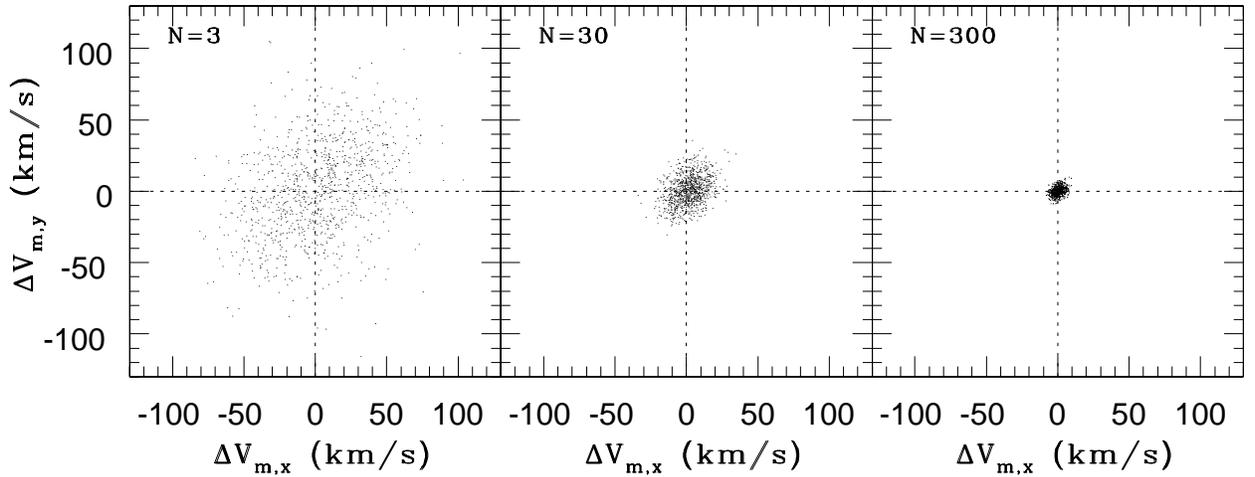}
\figcaption{
The same as Fig. 2 but for the fiducial LMC model with  $N=3$
(left), 30 (middle), and 300 (right).
\label{fig-3}}
\end{figure}

\newpage

\begin{figure}
\epsscale{0.8}
\plotone{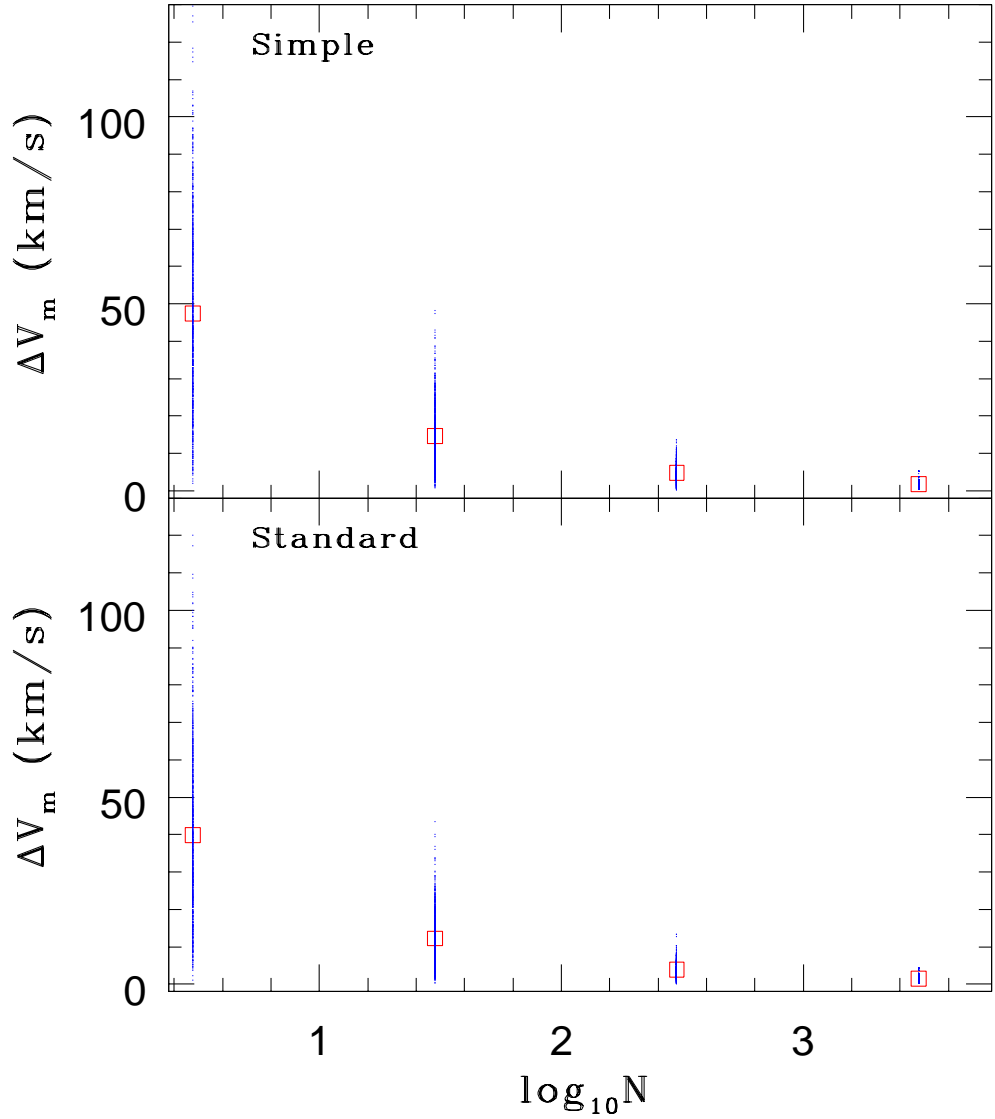}
\figcaption{
The distribution of $\Delta{\rm V}_{\rm m}$ at each $N$ (=3, 30, 300, and 3000)
in the simple (upper) and standard methods (lower). Each blue dot represents
the result of each sample. The red open squares represent the mean values  of
$\Delta{\rm V}_{\rm m}$ for 1000 smaples at each $N$. 
\label{fig-4}}
\end{figure}

\end{document}